\documentclass{ws-ijmpa}
\usepackage[super,compress]{cite}

\usepackage[T1]{fontenc}
\usepackage[english]{babel}
\usepackage{lmodern}
\usepackage{amsmath,bm,bbm}
\usepackage{verbatim}
\usepackage{csquotes}
\usepackage{simplewick}
\usepackage{appendix}
\usepackage[dvipsnames,svgnames]{xcolor}
\usepackage[normalem]{ulem}
\usepackage{hyperref}
\usepackage{cleveref}

\usepackage[framemethod=default]{mdframed}
\newmdenv[skipabove=7pt,
skipbelow=7pt,
rightline=false,
leftline=false,
topline=false,
bottomline=false,
backgroundcolor=gray!10,
linecolor=gray,
innerleftmargin=5pt,
innerrightmargin=5pt,
innertopmargin=5pt,
innerbottommargin=5pt,
leftmargin=0cm,
rightmargin=0cm,
linewidth=4pt]{eBox}

\newcommand{\md}{\mathrm{d}}

\begin{document}
\markboth{A. Ganz, P. Martens, S. Mukohyama \& R. Namba}{Bispectrum from inflation/bouncing Universe in VCDM}

% \preprint{YITP-24-72, IPMU24-0027, RIKEN-iTHEMS-Report-xx}
\def\preprintnumbers#1{\par
	\vspace*{8pt}
	{\authorfont{\leftskip18pt\rightskip\leftskip
	\noindent{\it Preprint numbers:}\/:\ #1\par}}\par}

\title{Bispectrum from inflation/bouncing Universe in VCDM}

\author{Alexander Ganz}

\address{Institute for Theoretical Physics, Leibniz University Hannover,\\
Appelstraße 2, 30167 Hannover, Germany\\
alexander.ganz@itp.uni-hannover.de}

\author{Paul Martens}

\address{Department of Physics, The Chinese University of Hong Kong (CUHK),\\ Shatin, New Territories, Hong Kong\\
paulmartens@cuhk.edu.hk}

\author{Shinji Mukohyama}
\address{
    Center for Gravitational Physics and Quantum Information (CGPQI),\\
    Yukawa Institute for Theoretical Physics (YITP),\\
    Kyoto University, 606-8502, Kyoto, Japan\\
    shinji.mukohyama@yukawa.kyoto-u.ac.jp
}
\address{
    Kavli Institute for the Physics and Mathematics of the Universe (WPI),\\
    The University of Tokyo, Kashiwa, Chiba 277-8583, Japan
}

\author{Ryo Namba}
\address{RIKEN Interdisciplinary Theoretical and Mathematical Sciences (iTHEMS),\\
Wako, Saitama 351-0198, Japan\\
ryo.namba@riken.jp}

% \author{Alexander Ganz}
% \email{alexander.ganz@itp.uni-hannover.de}
% \affiliation{Institute for Theoretical Physics, Leibniz University Hannover,
% Appelstraße 2, 30167 Hannover, Germany}
% \affiliation{Faculty of Physics, Astronomy and Applied Computer Science, Jagiellonian University, 30-348 Krakow, Poland}

% \author{Paul Martens}
% \email{paul.martens@yukawa.kyoto-u.ac.jp}
% \affiliation{Chinese University of Hong Kong (CUHK), Shatin, New Territories, Hong Kong}

% \author{Shinji Mukohyama}
% \email{shinji.mukohyama@yukawa.kyoto-u.ac.jp}
% \affiliation{Center for Gravitational Physics and Quantum Information (CGPQI),
% Yukawa Institute for Theoretical Physics (YITP),
% Kyoto University, 606-8502, Kyoto, Japan}
% \affiliation{Kavli Institute for the Physics and Mathematics of the Universe (WPI),
% The University of Tokyo, Kashiwa, Chiba 277-8583, Japan}

% \author{Ryo Namba}
% \email{ryo.namba@riken.jp}
% \affiliation{RIKEN Interdisciplinary Theoretical and Mathematical Sciences (iTHEMS), Wako, Saitama 351-0198, Japan}
% \date{\today}

\maketitle

\begin{abstract}
    We discuss the non-linear interactions within the VCDM model, a type II minimally modified gravity model with the same number of degrees of freedom as in General Relativity but not connected to the latter by field redefinitions. During an inflationary phase in the early universe, if the VCDM potential does not modify the slow-roll behavior of the inflaton field, we recover, up to the leading order, the standard results for the bispectrum in slow-roll inflation. On the other hand, if the VCDM potential becomes dominant, the interactions can strongly deviate and even violate the Maldacena's consistency relation for the local non-Gaussianities. Furthermore, we apply the formalism to the recently introduced bouncing model in VCDM, and show that the bispectrum still respects current observational constraints. Future measurements on non-Gaussianities of the local type should provide the test ground for the model's validity.

    \keywords{bispectrum; minimally modified gravity; VCDM; inflation; bouncing Universe.}

    \preprintnumbers{YITP-24-72, IPMU24-0027}
\end{abstract}

% \ccode{PACS numbers:}

% \maketitle

\section{Introduction}

While General Relativity (GR) has so far proven very successful in describing our Universe, testing potential deviations from it remains crucial in order to account for the source of the accelerated expansion (dark energy), the observational evidence of additional matter-like content in the Universe (dark matter) and the cosmic behavior in the primordial Universe, as well as to investigate the quantum nature of gravity in the UV sector.
In recent years, a new class of modified gravity models, called \enquote{Minimally Modified Gravity} (MMG), are receiving an increased attention \cite{Lin:2018mip,Iyonaga:2018vnu,Gao:2019twq,Lin:2017oow,Mukohyama:2019unx,DeFelice:2020eju,Yao:2020tur,Aoki:2021zuy,Ganz:2022iiv}. These theories modify GR at low-energy, while not containing extra local degrees of freedom besides the standard two tensor degrees of freedom from GR.
They therefore avoid the common pitfalls that the presence of additional degrees of freedom may entice, such as ghosts or gradient instabilities.
In this work, we consider VCDM \cite{DeFelice:2020eju}, a specific class of MMG theories.
According to the classification introduced in Ref.~[\citen{Aoki:2018brq}], VCDM is a type-II theory, meaning that it does not admit an Einstein frame.
Moreover, unlike GR, VCDM is not invariant under full spacetime diffeomorphism: spacetime is foliated, and the foliation acquires a physical meaning.
Recent works have shown VCDM to be a fruitful framework within which we can approach cosmological questions, such as tensions in late-time cosmology \cite{DeFelice:2020prd,DeFelice:2020cpt,Akarsu:2024qsi} and, as this paper focuses on, a bouncing Universe as an early-Universe scenario \cite{Ganz:2022zgs}.

In parallel, modified gravity theories have provided a rich platform for describing the very early Universe and formulate the successful inflationary paradigm.
Higher order correlation functions, such as the scalar perturbation bispectrum and trispectrum, have become crucial tools to test and discriminate between different candidate models \cite{Chen:2006nt,Seery:2005wm,Akama:2019qeh,Creminelli:2005hu}.
In the standard canonical single-field slow-roll inflationary model, the bispectrum is slow-roll suppressed and, therefore, still beyond the reach of current experimental constraints from the cosmic microwave background (CMB) \cite{Planck:2019kim}.
Furthermore, it must satisfy the Maldacena's consistency relation \cite{Maldacena:2002vr,Creminelli:2004yq}.
The latter relates the squeezed limit of the non-Gaussianities to the spectral index of the power spectrum, as a consequence of the fact that long wavelength modes can be regarded as a global part of diffeomorphism and can be absorbed by spatial rescaling.
Thus, the detection of large non-Gaussianities of local type would provide a powerful smoking gun for deviations from single-scalar-field slow-roll models.

In multi-field inflation \cite{Rigopoulos:2005ae,Rigopoulos:2005us}, in the curvaton scenarios \cite{Lyth:2002my,Enqvist:2005pg}, or in alternatives to inflation like bouncing models \cite{Cai:2009fn,Koyama:2007if}, the bispectrum can be significantly enhanced and can be used to constrain the parameter space. Indeed, common matter bouncing models with small scalar-mode sound speed $c_s \ll 1$ predict a large bispectral non-linearity $f_{\mathrm{NL}} \gg 1$ , which are in conflict with the current observational constraints and has led to the establishment of strong no-go theorems for single-scalar-field matter bounces \cite{Quintin:2015rta,Li:2016xjb,Akama:2019qeh}.

In the present study, we exhibit the scalar cubic interaction of the VCDM model to investigate the impact of the non-dynamical scalar field in the model on higher-order correlation functions.
The matter sector is simply given here by a $k$-essence model as a leading, illustrative example.
In general, the cubic interaction can be highly involved due to the non-local interaction.
Therefore, similarly to the analysis in Ref.~[\citen{Bartolo:2021wpt}] for the Cuscuton model, we consider the case where the energy density of the matter sector is subdominant $\alpha \equiv (\rho+p)/H^2 \ll 1$, where $\rho$, $p$ and $H$ are the matter energy density, its pressure and the Hubble rate, respectively, and where the reduced Planck mass is normalized to unity. This parameter regime is relevant for dark energy or inflation, or scenarios where the VCDM model dominates the background.

Firstly, we show that as long as the VCDM model does not spoil the slow-roll behavior of the inflaton we can recover the usual results of GR with $k$-essence.
By relaxing these conditions, it is possible to construct inflationary models which fulfill the standard constraints at linear order while leading to large local non-Gaussianities, violating the Maldacena's consistency relation.
Secondly, we apply our analysis to the recently introduced bouncing model \cite{Ganz:2022zgs}, and we discuss how the bispectrum can be used to constrain the parameter space of this model.

The structure of the paper is as follows. \Cref{sec:Recap} briefly recapitulates the construction of the VCDM model and gives a short review of the linear perturbations around the cosmological background. In \Cref{sec:Cubic_Interactions}, we derive the cubic scalar action and discuss its properties in the case of the standard slow-roll inflation.
Using the previous results, the applications to the bouncing model are discussed in \Cref{sec:Bouncing_model}. \Cref{sec:Discussion} is finally devoted to discussing our results.
Throughout the paper we work in units of the reduced Planck mass, \textit{i.e.} $M_{\text{Pl}}^2=(8\pi G)^{-1}=1$, and with a mostly positive signature of the metric.

\section{Recapitulation of VCDM}
\label{sec:Recap}

While the detailed construction and formulation of VCDM was exposed in Ref.~[\citen{DeFelice:2020eju}], we here review the model's definition and its features that are relevant for the current work.
The VCDM model is given by the action
\begin{align}
    S_\mathrm{VCDM} = \frac{1}{2} \int \md^4x N \sqrt{\gamma}
    \Big[ &
    R + K_{ij} K^{ij} - K^2 - 2 V(\varphi) \nonumber \\
    & - 2 \, \frac{\lambda^i}{N} \, \partial_i \varphi - \frac{3}{2} \, \lambda_0^2 - 2 \lambda_0 \left( K + \varphi \right)
    \Big]~,
    \label{action_mmg}
\end{align}
with $K_{ij}$ being the extrinsic curvature, $K$ its trace, $R$ the Ricci-scalar of the 3-dimensional intrinsic curvature associated with $\gamma_{ij}$, $N$ and $N^k$ respectively the lapse and the shift vector, $\lambda_0$ and $\lambda^i$ fields playing a role of Lagrange multipliers, and $\varphi$ an auxiliary scalar field. Due to the presence of constraint equations, the VCDM model propagates only two tensor degrees of freedom as in GR at the cost of dismissing temporal diffeomorphism invariance. For the matter sector we consider a $k$-essence model with the scalar field $\chi$, that is
\begin{subequations}
    \begin{align}
        S_\mathrm{matter} & = \int \md^4x N \sqrt{\gamma} \, P(X,\chi) ~, \\
        X & \equiv - \frac{1}{2} \, g^{\mu\nu} \partial_\mu \chi \, \partial_\nu \chi
        = \frac{1}{2} \left[ \left( \partial_\perp \chi \right)^2 - \gamma^{ij} \partial_i \chi \, \partial_j \chi \right]~,
    \end{align}
\end{subequations}
with $\partial_\perp \chi = \frac{1}{N}(\dot \chi - N^k \partial_k \chi)$, where $N^k$ is the shift vector.

\subsection{FLRW background}
For a homogeneous and isotropic background with no intrinsic curvature, we generically have
\begin{equation}
    \begin{gathered}
        \label{BG_metric}
        N = \bar{N}(t) = 1 \; , \qquad
        N^i = 0 \; , \qquad
        \gamma_{ij} = a^2(t) \, \delta_{ij} \; , \qquad
        \varphi = \phi(t) \; , \\
        \lambda^i = 0 \; , \qquad
        \lambda_0 = \bar\lambda(t) \; , \qquad
        \chi = \bar\chi(t) \; .
    \end{gathered}
\end{equation}
The equations of motion are then given by variation of the background action with respect to $\bar{N}$ (later set to $1$), $\bar\lambda$, $\phi$, $a$ and $\bar\chi$, respectively reading,
\begin{align}
    \label{eq:3}
    3 H^2 & = V + \phi \bar \lambda + \frac{3}{4} \, \bar \lambda^2 + \rho \; , \\
    0 & = 3 H + \phi + \frac{3}{2} \, \bar\lambda \; , \\
    0 & = \bar\lambda + V_\varphi \; , \\
    2 \, \dot H + 3 H^2 & = V + \phi \bar\lambda + \frac{3}{4} \, \bar\lambda^2 - \dot \lambda - p \; , \\
    0 & = \dot \rho + 3 H (\rho + p) \; ,
\end{align}
where overdot denotes derivative with respect to $t$, $H \equiv \dot a / a$, and
\begin{align}
    \rho = 2 X P_X - P \; , \qquad p = P \; ,
\end{align}
are all evaluated at the background values. For later purposes, we also define the parameters
\begin{equation}
    \begin{gathered}
        \epsilon = - \frac{\dot H}{ H^2}~, \qquad  \eta= \frac{\dot \epsilon}{\epsilon H }~, \qquad \alpha = \frac{\rho + p}{H^2}~, \qquad \eta_\alpha= \frac{\dot \alpha}{\alpha H}~, \\
        c_s^2 = \frac{P_X}{P_X + 2 X P_{XX}}~, \qquad s = \frac{\dot c_s}{H c_s}~.
    \end{gathered}
\end{equation}
Note that, in inflationary models of a single scalar field with GR, some of the above parameters are related to each other as $\epsilon = \alpha/2$ and $\eta = \eta_\alpha$. In our current case, the VCDM model admits freedom to liberate these relations, and thus the parameters defined above are mutually independent, subject to the background dynamics.

\subsection{Linear perturbations}
At the linear level, we can parameterize the scalar and tensor perturbations as
\begin{equation}
    \label{eq:11}
    \begin{gathered}
        N = 1  + \delta N  \; , \qquad
        N^i = \frac{1}{a(t)}  \delta^{ij}\partial_j \beta \; , \\
        \gamma_{ij} = a^2(t) \, {\rm e}^{2 \zeta} \left[ \delta_{ij} + h_{ij} + \frac{1}{2} \, h_{ik} h_{kj}  \right] \; , \\
        \varphi = \phi(t) + \delta\varphi(t, \bm{x}) \; , \qquad
        \chi = \bar\chi(t) + \delta\chi(t, \bm{x}) \; , \\
        \lambda_0 = \bar\lambda(t) + \delta\lambda_0(t, \bm{x}) \; , \qquad
        \lambda^i = \frac{1}{a^2} \delta^{ij}  \partial_j \delta\lambda_s(t, \bm{x})  \; ,
    \end{gathered}
\end{equation}
As an abuse of notation we hereafter drop the bar on the background quantities. Note that the VCDM model is not invariant under the full diffeomorphism invariance but instead only under spatial transformations
\begin{align}
    x^k \rightarrow x^k + \xi^k~,
\end{align}
which has been used to gauge fix the off-diagonal scalar metric perturbation.
All the vector modes are non-dynamical thanks to the spatial diffeomorphism and the associated constraints, and since they are decoupled from the scalar and tensor sectors at the linear level due to the background symmetry, they have no contribution up to cubic orders in the action, \textit{i.e.}~up to what we consider in the following analysis. Therefore we omit them from the beginning of our consideration.

Since the VCDM model does not introduce an additional dynamical scalar degree of freedom, there is only one scalar degree of freedom coming from the matter sector. It is convenient to work with the curvature perturbation
\begin{align}
    \mathcal{R} = \zeta - \frac{H}{\dot \chi} \delta \chi ~,
    \label{eq:Curvature_Linear}
\end{align}
which is gauge invariant under the full coordinate transformation $x^\mu \rightarrow x^\mu + \xi^\mu$ at the linear level. We work here with this gauge-invariant quantity, but, at higher orders, the definition of the curvature perturbation $\mathcal{R}$ also requires higher order corrections to keep it gauge-invariant at the corresponding order.
This is discussed in section \ref{subsec:Higher_Order_Curvature}.

The tensor modes can be decomposed into the two usual polarization modes as
\begin{align}
     h_{ij}(\tau,\mathbf{x}) = \sum_\sigma \int  \frac{\md^3k}{(2\pi)^{3}} e^{i \mathbf{k}\cdot \mathbf{x}} \epsilon^{\sigma}_{ij} h_\sigma(\tau,\mathbf{k})~,
\end{align}
where $\tau$ is the conformal time and  $\epsilon^\sigma_{ij}$ the  polarization tensor. The second order action for the tensor modes can then be read as
\begin{align}
    S_T^{(2)} = \frac{1}{8} \int \md \tau \, \md^3k \, a^2 \left[ h_\sigma^{\prime 2} - k^2 h_\sigma^2 \right]~,
\end{align}
where prime denotes derivative with respect to $\tau$.
The tensor quadratic action is the same as the one in GR, which is no surprise as the VCDM action \eqref{action_mmg} is constructed in such a way that only the scalar sector is modified at the linear order.

After introducing the curvature perturbation in \cref{eq:Curvature_Linear}, we integrate out the other non-dynamical scalar perturbation variables as
\begin{align}
    & \delta \lambda_0 = \delta \lambda_s = \delta \varphi = 0~, \\
    & \beta = \frac{3 \alpha a H^2 }{2 k^2 \dot \chi} \delta \chi ~, \\
    & \delta N = - \frac{2 c_s^2 k^2  }{ a^2 H^2 \alpha} \zeta - \frac{H^3 \alpha  (3 + 3 c_s^2 + \eta_\alpha - 2 \epsilon)-\dot \chi P_\chi }{H^2 \alpha \dot \chi} \delta \chi + \frac{\delta \dot \chi}{\dot \chi}~, \\
    & \delta \chi = - \frac{\dot \chi}{H} \frac{4 k^2 c_s^2  \mathcal{R} + 2 a^2 \alpha  H \dot{\mathcal{R}} }{4 c_s^2 k^2 + a^2 H^2 \alpha ( 6 c_s^2 + \alpha - 2 \epsilon) } \;. \label{eq:delta_chi_sol}
\end{align}
The second order action in conformal time is then given by
\begin{align}
    S_S^{(2)} = \int \md \tau \, \md^3k \, \frac{z^2(\tau,k)}{2} \left[ \mathcal{R}^{\prime 2} - c_\mathcal{R}^2(\tau,k) \, k^2 \mathcal{R}^2\right]~,
\end{align}
where
\begin{align}
    z^2(\tau,k) ={}& 2 a^2  \alpha \, \frac{2 k^2 + 3 \alpha \mathcal{H}^2  }{4 c_s^2 k^2 + \alpha \mathcal{H}^2 (6 c_s^2 + \alpha - 2 \epsilon) }  ~,\\
    c_\mathcal{R}^2(\tau,k) ={}
        &
        \frac{1}{(2 k^2 + 3 \alpha \mathcal{H}^2 ) (4 c_s^2 k^2 + \alpha \mathcal{H}^2 ( 6 c_s^2 + \alpha - 2 \epsilon ))}
        \Big[
            8 c_s^4 k^4 \nonumber\\
            & + 2 c_s^2 k^2 \mathcal{H}^2 \big(\alpha^2 + 2 \alpha (-3 + 3 c_s^2 + 2 s - 2 \eta_\alpha + \epsilon ) \nonumber\\
            & \phantom{+ 2 c_s^2 k^2 \mathcal{H}^2 \big(}\, + 4 \epsilon (3 + 3 c_s^2 + \eta_\alpha - 2 s + \eta-2 \epsilon) \big) \nonumber\\
            & + \alpha \mathcal{H}^4 \big(36 c_s^4 \epsilon - (\alpha- 2\epsilon)^2 +6 c_s^2 \alpha   (-1  + 2 s - \eta_\alpha  + \epsilon ) \nonumber\\
            & \phantom{+ \alpha \mathcal{H}^4 \big(}\, + 12 c_s^2 \epsilon (1- 2 s - \epsilon + \eta )  \big)
        \Big]~,
\end{align}
with the conformal Hubble parameter $\mathcal{H}= a H$. Note that, in the case of a GR background, we would have additional relations $\alpha = 2 \epsilon$ and $\eta_\alpha = \eta$, with which the usual results of the $k$-essence model would be recovered at linear order, \textit{i.e.}~$c_\mathcal{R}^2= c_s^2$ and $z^2 = 2 a^2 \epsilon/c_s^2$.
Perturbations in VCDM have been studied in Refs.~[\citen{DeFelice:2020eju,DeFelice:2020prd,Sangtawee:2021mhz,Ganz:2022iiv}], and in particular for the same model (but without a $\chi$ dependence in $P(X,\chi)$) in Ref.~[\citen{Ganz:2022zgs}].

\section{Cubic interactions}
\label{sec:Cubic_Interactions}

The main goal of this paper is to test the VCDM model \eqref{action_mmg} by using the bispectrum of the curvature perturbations.
This section is devoted to the derivation of the cubic interaction terms, necessary to obtain the latter.
Observable gauge-invariant variables need to be constructed order by order, which we first do in \cref{subsec:Higher_Order_Curvature}, before expanding the Lagrangian up to the cubic order under a simplifying assumption in the subsequent subsections.

\subsection{Higher-order \texorpdfstring{$4$-D}{4-D} gauge-invariant curvature perturbation}
\label{subsec:Higher_Order_Curvature}
Using the active approach, the metric transforms at second order as \cite{Malik:2008im}
\begin{align}
    \tilde g_{\mu\nu} = e^{\mathcal{L}_\xi} g_{\mu\nu} = g_{\mu\nu} + \mathcal{L}_\xi g_{\mu\nu} + \frac{1}{2} \mathcal{L}_\xi^2 g_{\mu\nu} + \dots \; .
\end{align}
Therefore, the linear expression for the curvature perturbation $\mathcal{R}$ defined in \cref{eq:Curvature_Linear} is no longer $4$-dimensional gauge-invariant at second order. Instead, following Ref.~[\citen{Lyth:2005du}], we can define the 4-dimensional gauge-invariant curvature perturbation as~%
\footnote{We only need to expand $\mathcal{R}$ up to the second order, since the third-order perturbation $\mathcal{R}_3$ would appear only linearly in the cubic action, which vanishes once the background equations of motion are imposed.}
\begin{align}
    \mathcal{R}=\mathcal{R}_1 + \frac{1}{2} \mathcal{R}_2~,
\end{align}
where the lower index here denotes the order of perturbation, and thus
\begin{subequations}
    \begin{align}
        \mathcal{R}_{1} &
        = \zeta_1 - \frac{H}{\dot \chi} \delta \chi_1~,\\
        \mathcal{R}_{2} &
        = \zeta_2 -  \frac{H}{\dot \chi} \delta \chi_2 + \frac{2 H}{\dot \chi^2} \delta \chi_1 \delta \dot \chi_1- 2 \frac{\delta \chi_1}{\dot \chi} \dot \zeta_1 - \left( \frac{H \delta \chi_1}{\dot \chi}\right)^2 \left( \frac{\ddot \chi}{H \dot \chi}- \frac{\dot H}{H^2}\right)~,
    \end{align}
\end{subequations}
where $\delta\chi$ is similarly expanded up to the second order.
Inverting  it in terms of $\zeta$ yields, up to second order,
\begin{align}
    \zeta &
        \equiv \zeta_1 + \frac{1}{2}\zeta_2 = \mathcal{R} + \frac{H}{\dot \chi} \delta \chi - \frac{H}{\dot \chi^2} \delta \chi \delta \dot \chi + \frac{\delta \chi}{\dot \chi} \frac{\md}{\md t} \left(\mathcal{R} + \frac{H}{\dot \chi}  \delta \chi \right) + \frac{H^2\delta \chi^2}{2\dot \chi^2} \left( \frac{\ddot \chi}{H \dot \chi}+ \epsilon \right) \nonumber  \\
    &
        = \mathcal{R} + \frac{H}{\dot \chi} \delta \chi - \frac{H^2}{2\dot \chi^2}  \left(\frac{\ddot \chi}{H \dot \chi} + \epsilon\right) \delta\chi^2 + \frac{\delta \chi \dot{\mathcal{R}}}{\dot \chi}~.
\end{align}
The part of Lagrangian quadratic in $\delta\chi$ and $\zeta$ has the form
\begin{align}
    \mathcal{L}^{(2)}_{\zeta} =  a^3 \Big[ &
        - \frac{1}{\dot \chi} \zeta \left(
            2 \frac{\partial^2}{a^2} - 3 \alpha H^2
        \right) \delta \dot \chi
        - \zeta \left(
            \frac{\partial^2}{a^2} +  \frac{2 c_s^2 }{H^2 \alpha }\frac{\partial^4}{a^4}
        \right) \zeta \nonumber\\
    &
        + \frac{\alpha H^2}{4 \dot \chi^2}  \delta \chi \left(
            2 \frac{\partial^2}{a^2}  - 3  H^2 (6 + 6 c_s^2 + \alpha + 2 \eta_\alpha - 4 \epsilon )
        \right) \delta \chi \nonumber\\
    &
        + \zeta \left(  \frac{2 \partial^2}{a^2 \dot \chi} \left( H (3 + 3 c_s^2 + \eta_\alpha - 2\epsilon ) - \frac{\dot \chi P_\chi }{H^2 \alpha }\right) + 3 P_\chi \right) \delta \chi \Big]~,
\end{align}
which after  substituting the non-linear definition for the curvature perturbation leads to
\begin{align}
    \mathcal{L}^{(2)}_{\zeta} =  \mathcal{L}^{(2)}_{\mathcal{R}} + \tilde{\mathcal{L}}^{(3)}~,
    \qquad\text{where}\qquad
    \mathcal{L}^{(2)}_{\mathcal{R}} \equiv
    \left.\mathcal{L}^{(2)}_{\zeta}\right|_{\zeta \to \mathcal{R} + H\delta\chi/\dot{\chi}}~,
\end{align}
and
\begin{align}
    \tilde{\mathcal{L}}^{(3)} \equiv & \,
    a^3 \left(
    \frac{\delta \chi \dot{\mathcal{R}}}{\dot \chi}
    - \frac{H^2}{2\dot \chi^2} \left( \frac{\ddot \chi}{H \dot \chi} + \epsilon \right) \delta \chi ^2
    \right)
    \nonumber\\
    & \times
    \bigg[
    - \frac{1}{\dot \chi}  \left( 2 \frac{\partial^2}{a^2} - 3 \alpha H^2 \right) \delta \dot \chi
    %\nonumber\\
    %&
    - 2 \left( \frac{\partial^2}{a^2} + \frac{2 c_s^2 }{H^2 \alpha} \frac{\partial^4}{a^4} \right) \left( \frac{H}{\dot\chi} \delta \chi + \mathcal{R}\right) \nonumber\\
    & \qquad
    + \left(  \frac{2 \partial^2}{a^2 \dot \chi} \left( H (3 + 3 c_s^2 + \eta_\alpha - 2\epsilon ) - \frac{\dot \chi P_\chi }{H^2 \alpha }\right) + 3 P_\chi \right) \delta \chi
    \bigg]~.
\end{align}
In general, we need to solve the perturbations of $\delta \chi$ up to quadratic order. However, as for the lapse and shift vector the prefactor of the quadratic contributions is proportional to the constraint equation and vanishes.

Finally, the cubic Lagrangian can be expressed as
  \begin{align}
   \mathcal{L}^{(3)}_{\mathcal{R}} = \left.\mathcal{L}^{(3)}_{\zeta}\right|_{\zeta \to \mathcal{R} + H\delta\chi/\dot{\chi}} + \tilde{\mathcal{L}}^{(3)}~, \qquad\text{where}\quad
   \mathcal{L}^{(3)}_{\zeta} = \mathcal{L}^{(3)}_{\rm VCDM} + \mathcal{L}^{(3)}_{\rm matter}~.
  \end{align}
Here, $\mathcal{L}_{\rm VCDM}^{(3)}$ and $\mathcal{L}_{\rm matter}^{(3)}$ are the cubic interactions of the gravity and matter sectors, respectively, that are present already in terms of $\zeta$ and other original variables. In these terms, we can simply replace $\zeta \to \mathcal{R} + H\delta\chi/\dot{\chi}$, as we are interested in up to the cubic action.
Let us note that, while $\mathcal{L}^{(3)}_{\mathcal{R}}$ appears to depend on $\mathcal{R}$ and $\delta\chi$ at this stage, the number of the true dynamical degree of freedom is only $1$, thanks to the additional constraints in the VCDM theory. Solving the constraint equations without any approximation is computationally hard, and hence in the following subsection we proceed the calculation under a relevant approximation, showing that the action indeed depends only on one variable, which we take to be $\mathcal{R}$, after solving the constraints.

\subsection{Expansion in \texorpdfstring{$\alpha$}{α}}
\label{sec:Expansion_in_alpha}
In general, cubic or higher order interactions are quite involved due to the elliptic equations that eliminate unwanted extra degrees of freedom. However, similarly to the analysis in Ref.~[\citen{Bartolo:2021wpt}], the calculations simplify drastically if $\alpha = (\rho + p)/H^2 \ll 1$. This approximation is valid in two distinct scenarios: either we have a nearly de-Sitter background with an equation of state $\omega \simeq -1$, which is relevant for inflation or dark energy, or during a period where the VCDM potential $V(\varphi)$ dominates the background evolution. The latter case becomes important for the second part of the paper where we apply the formalism to the bouncing model introduced in Ref.~[\citen{Ganz:2022zgs}].

Considering the linear solution of  $\delta \chi$ in terms of the curvature perturbation $\mathcal{R}$ (\cref{eq:delta_chi_sol}),
the expression is regular, and there are no divergences as long as $6 c_s^2 + \alpha - 2 \epsilon >0$, which we assume from here on.  Therefore, we expand it in orders of $\alpha \mathcal{H}^2/k^2$, which, inside or around horizon crossing $k/\mathcal{H} \lesssim \mathcal{O}(1)$, just corresponds to an expansion in orders of $\alpha $ similar to the slow-roll expansion in inflation. Under this approximation we have
\begin{align}
    \delta \chi = - \frac{\dot \chi}{H} \mathcal{R} + \frac{\dot \chi}{H} \, \alpha \, \frac{(3 c_s^2 - \epsilon) a^2 H^2 \mathcal{R} - a^2 H \dot{\mathcal{R}}  }{2 k^2 c_s^2}  + \mathcal{O}(\alpha^2)~.
\end{align}
As for $\zeta$, we obtain
\begin{align}
    \zeta = \mathcal{R} + \frac{H}{\dot \chi} \, \delta \chi = a^2 \alpha \, \frac{(3 c_s^2 - \epsilon)  H^2 \mathcal{R} -  H \dot{\mathcal{R}}}{2 c_s^2 k^2}~,
\end{align}
and observe that $\zeta$ is of order $\mathcal{O}(\alpha)$.
Similarly, we get
\begin{align}
    \delta N =& - \frac{2 c_s^2 k^2  }{ a^2 H^2 \alpha} \, \zeta - \frac{H^3 \alpha  (3 + 3 c_s^2 + \eta_\alpha - 2 \epsilon)-\dot \chi P_\chi }{H^2 \alpha \dot \chi} \, \delta \chi + \frac{\delta \dot \chi}{\dot \chi}~,\nonumber \\
    \simeq & - \frac{ \alpha a^2}{2 k^2 c_s^2} \ddot{\mathcal{R}} + \frac{\alpha (3 c_s^2 - 2 + 2 s -\eta_\alpha) a^2 H}{2 k^2 c_s^2} \dot{\mathcal{R}} \nonumber\\
    & - \frac{\alpha (k^2 c_s^2 + ( - 3 c_s^2 (2 + \eta_\alpha - 2 \epsilon ) +\epsilon (2 - 2 s + \eta_\alpha - 2 \epsilon + \eta ) ) a^2 H^2 )}{2 k^2 c_s^2} \mathcal{R}~, \\
    \beta =& \frac{3 \alpha a H^2 }{2 k^2 \dot \chi} \, \delta \chi \simeq - \frac{3 \alpha a H}{2 k^2} \mathcal{R}~,
\end{align}
where we have assumed that $\eta_\alpha =\mathcal{O}(1)$ or smaller and does not diverge in the limit $\alpha \rightarrow 0$. Therefore, $\beta = \mathcal{O}(\alpha)$ and $\delta N = \mathcal{O}(\alpha)$.

Since $\beta,\,\delta N$ and $\zeta$ are all of order $\mathcal{O}(\alpha)$, we conclude that $\mathcal{L}_\mathrm{VCDM}^{(3)}=\mathcal{O}(\alpha^3) $, and we can approximate the cubic perturbation of the Lagrangian as
\begin{align}
 \mathcal{L}^{(3)}_{\mathcal{R}}
 = \mathcal{L}^{(3)}_\mathrm{VCDM} +
        \mathcal{L}_\mathrm{matter}^{(3)} + \tilde{\mathcal{L}}^{(3)}
 \simeq \mathcal{L}_\mathrm{matter}^{(3)} + \tilde{\mathcal{L}}^{(3)} \; .
\end{align}
After some straightforward calculations, we find that the leading order terms read
\begin{align}
    \mathcal{L}^{(3)}_\mathrm{matter} \simeq{} a^3 \bigg[ & \frac{H^3 \alpha}{c_s^2 \dot \chi^3} \left( 3 + 3 c_s^2 + \eta_\alpha -2 \epsilon - (1+c_s^2) \frac{\dot \chi P_\chi}{H^3 \alpha } \right) \delta \chi \, \frac{(\partial_i \delta \chi)^2}{a^2} + \frac{2 \ell }{\dot \chi^3} \, \delta \dot \chi^3 \nonumber\\
    & - \frac{1}{2 c_s^2 \dot \chi^3} \Big( H^3 \alpha ( 6 + 2 s + \eta_\alpha -2 \epsilon) - 2 \dot \chi  P_\chi \nonumber\\
    & \phantom{- \frac{1}{2 c_s^2 \dot \chi^3}\Big(}\, + \frac{12 c_s^2 \ell }{H^2 \alpha }  (H^3 \alpha (3 + \eta_\alpha -2 \epsilon) - \dot \chi P_\chi ) \Big)  \delta \chi \, \delta \dot \chi^2 \nonumber\\
    & + \frac{\dot \chi P_{X\chi\chi}}{2} \, \delta \dot \chi \, \delta \chi^2  + \frac{c_s^2 -1}{2 c_s^2 \dot \chi^3} H^2 \alpha \, \delta \dot \chi \, \frac{(\partial_k \delta \chi)^2}{a^2} + \frac{P_{\chi\chi\chi}}{6} \, \delta \chi^3 \bigg]~,
\end{align}
where we have defined
\begin{align}
    \ell = \frac{1 - c_s^2}{4 c_s^2} H^2 \alpha + \frac{\dot \chi^6 P_{XXX}}{12}~.
\end{align}
On the other hand, the contributions from the non-linear definition of the curvature perturbation can be expressed as
\begin{align}
    \tilde{\mathcal{L}}^{(3)} \simeq{}& \left( \frac{ \mathcal{R}\dot{\mathcal{R}}}{H} + \frac{1}{2} \left( \frac{\ddot \chi}{H \dot \chi} + \epsilon \right)  \mathcal{R}^2 \right)
    \nonumber\\
    & \times
    \left(   a^3 \alpha \epsilon \frac{H^2 }{c_s^2}  (3 + 3 c_s^2 - 2 s + \eta_\alpha - 2 \epsilon + \eta ) \mathcal{R} +  \frac{\delta \mathcal{L}}{\delta \mathcal{R}}  \right)~,
\end{align}
where the linear equations of motion read
\begin{align}
    \frac{\delta \mathcal{L}}{\delta \mathcal{R}} \simeq a^3 \frac{\alpha}{c_s^2} \left( \ddot{\mathcal{R}} + H (3 - 2 s+ \eta_\alpha ) \dot{\mathcal{R}} - c_s^2 \frac{\partial^2}{a^2} \mathcal{R} \right)~,
\end{align}
which would vanish on-shell.

\subsection{Inflation scenarios}
Let us first focus on the inflationary case.
Depending on whether the VCDM potential dominates the background and on the evolution of the dynamical scalar field, very different predictions can be drawn. We now separately consider an inflationary scenario of slow-roll type and a toy model with a constant equation of state $w$ where the background evolution is dominated by the VCDM potential.

\subsubsection{Slow-roll evolution}
The equation of motion for the scalar field can be expressed as
\begin{align}
    \ddot \chi = H (3 + \eta_\alpha - 2 \epsilon) \dot \chi - \frac{P_\chi \dot \chi^2}{H^2 \alpha} \;.
\end{align}
Therefore, if we want to consider a regime where the evolution of the inflaton still obeys the slow-roll conditions, \textit{i.e.} $\epsilon, s, \eta, \alpha, \eta_\alpha \ll 1$ and $\ddot \chi \ll H \dot \chi$, we get the condition
\begin{align}
     \alpha \simeq  \frac{P_\chi \dot \chi}{3 H^3}\,.
\end{align}
Note that we do not assume that the potential $V(\phi)$ of VCDM is dominant or not, but instead only require that the evolution of the scalar field is not significantly modified.
In this case, at leading order, the contributions of $\tilde{\mathcal{L}}^{(3)}$ simplify down to
\begin{align}
    \tilde{\mathcal{L}}^{(3)} \simeq \frac{\mathcal{R} \dot{\mathcal{R}}}{H} \frac{\delta \mathcal{L}}{\delta\mathcal{R}}~,
\end{align}
which does not contribute to the bispectrum since the curvature perturbation is conserved outside the horizon.
Similarly, using that $\delta \chi \simeq - \frac{\dot \chi}{H} \mathcal{R}$, we can simplify the cubic contributions from the matter sector as
\begin{align}
    \mathcal{L}^{(3)}_\mathrm{matter}
    \simeq & a^3 \alpha \Big[ - \frac{c_s^2 -1}{2 c_s^2  H} \dot{\mathcal{R}} \frac{(\partial_k \mathcal{R})^2}{a^2}   - \frac{2 \ell}{H^3 \alpha} \dot{\mathcal{R}}^3  \Big]~.
\end{align}
By using the linear equation of motion and performing integration by parts, we can rewrite it as
\begin{align}
    \mathcal{L}^{(3)}_\mathrm{matter} \simeq a^3 \Big[
        & -  \left(  \frac{H^2 \alpha  (c_s^2-1)}{2 c_s^4}  + 2 \ell \right) \frac{\dot{\mathcal{R}}^3}{H^3} +  \frac{3 \alpha}{2 c_s^4} (c_s^2-1) \dot{\mathcal{R}}^2 \mathcal{R} \nonumber\\
        & - \frac{\alpha}{2c_s^2} (c_s^2-1) \frac{(\partial_k \mathcal{R})^2}{a^2} \mathcal{R} \Big] - \frac{1}{H c_s^2} (c_s^2-1) \mathcal{R} \dot{\mathcal{R}} \, \frac{\delta \mathcal{L}}{\delta \mathcal{R}}~.
\end{align}
This is consistent with the standard results from $k$-essence at leading order in slow-roll \cite{Seery:2005wm,Chen:2006nt}. Therefore, as long as the VCDM potential does not break the standard slow-roll assumption during inflation the results coincide with GR and $k$-essence by taking the GR limit $\alpha=2 \epsilon$ up-to-leading order in slow-roll parameters.

\subsubsection{Non-standard evolution}
Due to the presence of the VCDM potential, it is possible to reconstruct essentially any desired background evolution \cite{DeFelice:2020eju,Ganz:2022zgs}. Therefore, by tuning the potential such that it dominates the background evolution we can break the standard assumptions on the scalar field evolution.
In what follows, we construct a toy model to see potentially new signatures in the bispectrum.
We adopt the approach of Ref.~[\citen{Ganz:2022zgs}] to reconstruct the VCDM potential.
For simplicity, we take a power-law ansatz, \textit{i.e.}
\begin{align}
    a \propto t^p~,
\end{align}
with $p> 1$ to realize $\ddot a>0$ and a $k$-essence fluid with a constant equation of state
\begin{align}
    P = P_0 X^{\frac{1+w}{2w}}~.
\end{align}
By tuning the power-law index $p$ and the equation of state $w$ we can fulfill the requirements on a nearly flat scalar power-spectrum and frozen curvature perturbations outside the horizon (see \cref{app:Inflation_Toy_Model} for more details).

The cubic action  differs from the slow-roll case since $\eta_\alpha = \mathcal{O}(1)$ and $1 >\epsilon > \frac{3}{4}$ (which results from the above requirements on $p$ and $w$) and is given by
\begin{align}
    \begin{split}
    \mathcal{L}^{(3)} \simeq{} & a^3 \Bigg\{
        \frac{1}{2} \frac{1}{c_s^2 \dot \chi^3} H^2 \alpha (c_s^2-1) \frac{(\partial_i \delta \chi)^2}{a^2} \, \delta \dot\chi - \frac{ c_s^2 - 1 }{6 c_s^4 \dot \chi^3}  \, \alpha H^2 \delta \dot \chi^3 \Bigg\} \\
        & +  \left( \frac{\mathcal{R}\dot{\mathcal{R}}}{\dot \chi} + \frac{1}{2}  (\epsilon - 3 c_s^2) \mathcal{R}^2
    \right) \frac{\delta \mathcal{L}}{\delta \mathcal{R}}
    \end{split} \nonumber \\
    \simeq{} &  - a^3 \alpha \Bigg\{
        \frac{c_s^2-1}{c_s^2}   \frac{1}{2} \frac{(\partial_i \mathcal{R})^2}{a^2} \left( (\epsilon - 3 c_s^2) \mathcal{R} + \frac{\dot{\mathcal{R}}}{H} \right) \nonumber\\
        & \phantom{- a^3 \alpha \Bigg\{}\, - \frac{ c_s^2 - 1 }{6 c_s^4}  \left( 3 (\epsilon - 3 c_s^2) \mathcal{R} \dot{\mathcal{R}}^2 + \frac{\dot{\mathcal{R}}^3}{H} \right)
    \Bigg\} \nonumber\\
    & + \left( \frac{\mathcal{R}\dot{\mathcal{R}}}{\dot \chi} + \frac{1}{2}  (\epsilon - 3 c_s^2) \mathcal{R}^2 \right) \frac{\delta \mathcal{L}}{\delta \mathcal{R}}~.
    \label{eq:Cubic_action_constant_eqs}
\end{align}
The details of the bispectrum are discussed in \Cref{sec:Bouncing_model}, where we cover the case of the bouncing model. For the current purposes, most interesting is the squeezed limit $k_1 \ll k_2\simeq k_3$, or local shape of the bispectrum, which is given by
\begin{align}
    f_\text{NL}^{loc} = \frac{5}{12 c_s^2} (\epsilon - 3 c_s^2)  = \frac{5}{16 w}  (1-3w) ~.
\end{align}
The toy model can thus provide significant local non-Gaussianities, especially for $c_s \ll 1$.
It violates the Maldacena's consistency relation \cite{Maldacena:2002vr,Creminelli:2004yq}, which can be related to the presence of the VCDM potential violating the single clock assumption.

\section{Bouncing model}
\label{sec:Bouncing_model}

\subsection{Recapitulation of the bouncing model}
The VCDM model has a rich structure that can accommodate otherwise difficult-to-realize cosmological scenarios, without introducing any instabilities.
As our leading scenario, let us briefly review the bouncing model constructed in Ref.~[\citen{Ganz:2022zgs}].
For the matter sector we consider a $k$-essence model with a constant equation of state $w$, which can be achieved by the functional form
\begin{align}
    P(X) = P_0 X^{\frac{1+w}{2w}}~.
\end{align}
By tuning the potential $V(\varphi)$, we can fix the background evolution such that the scale factor reads
\begin{equation}
    \label{eq:Scale_factor}
    a (\tau) = a_0 \left( \frac{\tau^2}{\tau_e^2} \right)^{\frac{n}{2}} \Theta(\tau_e-\tau) + a_1 \left[ 1+ \left( \frac{\tau}{\tau_B} \right)^2 \right]^{\frac{1}{3w+1}} \Theta(\tau-\tau_e)\;,
\end{equation}
with $a_1=a_0 \left(1+(\tau_e/\tau_B)^2\right)^{-1/(3w+1)}$.
During the contraction phase $\tau \ll \tau_e $, the VCDM potential dominates, which leads to a subdominant energy density of the scalar field $\alpha \ll 1$.
By setting the power-law coefficient $n$ of the scale factor during the contraction phase as
\begin{align}
    n =   \frac{5-n_s}{3w -1}~,
\end{align}
we recover the spectral index $n_s$ for the scalar perturbations that leave the horizon before $\tau \ll \tau_e$.
Outside the horizon, the scalar modes freeze until $\alpha \mathcal{H}^2 \sim k^2$, when the modes briefly drop before freezing again. The bounce itself does not impact the curvature perturbation (see Ref.~[\citen{Ganz:2022zgs}] for more details).
The scalar power spectrum for the modes which leave the horizon at $\tau \ll \tau_e$ can be parameterized as
\begin{align}
    P_\mathcal{R} = \frac{1}{ 4 f \pi^2 z_{\star}^2 \tau_\star^2 c_s^3 }~,
\end{align}
where the $\star$ denotes the evaluation at horizon crossing and $f=\mathcal{O}(1)$ is denoting the drop of the scalar power spectrum at $\alpha \mathcal{H}^2 \sim k^2$, and can be evaluated numerically. As it is of order one, we omit it from our discussion from now on.

The tensor modes, on the other hand, behave like many other bouncing scenarios and grow outside the horizon during the contraction phase. The spectral index for the modes leaving during $\tau \ll \tau_e$ is given by $n_t = 4 -2 n$ and, therefore, for these modes the tensor-to-scalar ratio is, in general, scale dependent
\begin{align}
    r = r_0 \left( k \tau_B \right)^{n_t - n_s -1}~.
\end{align}
After the bounce $r_0 \gg 1 $. We therefore need to require that $n_t - n_s - 1 > 0$, which implies $w = c_s^2 > 1$, in order to fulfill the CMB constraints on the tensor-to-scalar ratio $r(k_\text{CMB}) < 0.056$ \cite{Planck:2018jri}.

\subsection{Scalar Bispectrum}
Using the in-in formalism the bispectrum for the curvature perturbation is given by
\begin{align}
    \langle \mathcal{R}^3 \rangle =& - i \int_{\tau_{in}}^\tau a \md \tau^\prime \langle [\mathcal{R}(\tau,\mathbf k_1) \mathcal{R}(\tau,\mathbf k_2) \mathcal{R}(\tau, \mathbf k_3), H_{int}(\tau^\prime) ] \rangle~.
\end{align}
As discussed earlier, during the contraction phase for $\tau \ll \tau_e $, the VCDM potential dominates the background evolution such that $\alpha \ll 1$. Therefore, at CMB scales, we can use the expansion in orders of $\alpha \mathcal{H}^2/k^2$ in \Cref{sec:Expansion_in_alpha}. Outside the horizon, the curvature perturbation is frozen, contrary to common bouncing scenarios. Therefore, it does not contribute significantly to the bispectrum. At $\alpha \mathcal{H}^2 \sim k^2$ the modes unfreeze for a short timescale and it could contribute to the bispectrum. However, since both the timescale and the change in the curvature perturbation are small we expect the contribution to be suppressed. Similarly, during the bounce itself, the curvature perturbation on CMB scales is frozen and does not impact the bispectrum.

The cubic action in this model under relevant approximations is given in \cref{eq:Cubic_action_constant_eqs}.
% \begin{align} % Since the foloowing exactly matches \cref{eq:Cubic_action_constant_eqs}, shall we omit it altogether and simply refer to the corresponding equation?
%     \mathcal{L}^{(3)}
%     \simeq{} &  - a^3 \alpha \Bigg\{
%         \frac{c_s^2-1}{c_s^2}   \frac{1}{2} \frac{(\partial_i \mathcal{R})^2}{a^2} \left( (\epsilon - 3 c_s^2) \mathcal{R} + \frac{\dot{\mathcal{R}}}{H} \right) \nonumber\\
%         & \phantom{- a^3 \alpha \Bigg\{}\, - \frac{ c_s^2 - 1 }{6 c_s^4}  \left( 3 (\epsilon - 3 c_s^2) \mathcal{R} \dot{\mathcal{R}}^2 + \frac{\dot{\mathcal{R}}^3}{H} \right)
%     \Bigg\} \nonumber\\
%    & + \left( \frac{\mathcal{R}\dot{\mathcal{R}}}{\dot \chi} + \frac{1}{2}  (\epsilon - 3 c_s^2) \mathcal{R}^2 \right) \frac{\delta \mathcal{L}}{\delta \mathcal{R}}~.
% \end{align}
Assuming that $n_s\simeq 1$ the curvature perturbation for $\tau \ll \tau_e$ can be written as
\begin{align}
    \mathcal{R}(\tau^\prime,\mathbf k_1) \simeq \mathcal{R}_0 \left(1+ i c_s k_1 \tau^\prime \right) e^{-i c_s k_1 \tau^\prime}~,
\end{align}
with
\begin{align}
    \mathcal{R}_0(k_1) = \frac{1}{\sqrt{2} \, (c_s k_1)^{3/2} z_{\star} \tau_{\star}}~.
\end{align}
Note that, while $\alpha \mathcal{H}^2/k^2 \ll 1$, the curvature perturbations in contraction phase behave as in standard inflation. Therefore, the calculations are close to those carried out in standard inflation scenarios. However, $\epsilon = \mathcal{O}(1)$ is here not a slow-roll parameter.
%%%%%%%%%%%%%%%%%
Finally, we obtain
\begin{align}
    \langle \mathcal{R}^3\rangle =&  (2\pi)^7 \delta^{(3)}\left( \mathbf{k}_1+\mathbf{k}_2 + \mathbf{k}_3 \right)  P_{\mathcal{R}}^2  \Pi_j \frac{1}{k_j^3} \sum_m \mathcal{A}_m~,
\end{align}
with the shape functions
\begin{align}
    \mathcal{A}_{\dot{\mathcal{R}}^3} ={}&  \frac{3(c_s^2-1)}{4c_s^2}  \frac{3(w+1)}{4}   \left[ \frac{k_1^2 k_2^2 k_3^2}{K^3} \right]~, \\
    % \mathcal{A}_{\dot{\mathcal{R}}^2 \mathcal{R}} ={}&  \frac{c_s^2-1}{c_s^2} \frac{\epsilon - 3c_s^2}{8} \left[  \frac{k_1^2 k_2^2 }{K} \left( 1 + \frac{k_3}{K} \right) + \frac{k_1^2 k_3^2}{K} \left( 1 + \frac{k_2}{K} \right) + \frac{k_2^2 k_3^2}{K} \left(1 + \frac{k_1}{K} \right) \right]~, \\
    \mathcal{A}_{\dot{\mathcal{R}}^2 \mathcal{R}} ={}&  \frac{c_s^2-1}{c_s^2} \frac{\epsilon - 3c_s^2}{8} \left[
          \frac{k_1^2 k_2^2}{K} \frac{K + k_3}{K}
        + \frac{k_1^2 k_3^2}{K} \frac{K + k_2}{K}
        + \frac{k_2^2 k_3^2}{K} \frac{K + k_1}{K}
    \right]~, \\
    \mathcal{A}_{\mathcal{R}(\partial \mathcal{R})^2} % =&  -\frac{(c_s^2-1) (\epsilon -3 c_s^2) }{8 c_s^2}  \left[ (\mathbf{k}_1 \cdot \mathbf{k}_2) \left( - K + \frac{ k_1 k_2 + k_1 k_3 + k_2 k_3}{K} + \frac{k_1 k_2 k_3}{K^2} \right) + \text{sym.} \right] \nonumber \\
    ={}& \frac{(c_s^2-1) (\epsilon-3c_s^2)}{8 c_s^2} \frac{1}{2} \sum_j k_j^2 \Big[ - K + \frac{1}{2K} \sum_{i\neq j} k_i k_j + \frac{k_1 k_2 k_3}{K} \Big]~, \\
    \mathcal{A}_{\delta L/\delta \mathcal{R}} ={}&  \frac{ \epsilon - 3c_s^2 }{8} \sum k_i^3~, \\
    \mathcal{A}_{\dot{\mathcal{R}} (\partial \mathcal{R})^2} ={}&  \frac{c_s^2-1}{8 c_s^2} \frac{3(w+1)}{4}  \left[ \frac{k_3^2}{K} (\mathbf{k}_1\cdot \mathbf{k}_2) \left( 1 + \frac{k_1 + k_2}{K} + \frac{2 k_1 k_2 }{K^2} \right) + \text{sym.}\right]~.
\end{align}
where we wrote $K \equiv k_1 + k_2 + k_3$ and ``sym.'' denotes the terms symmetrized among the three wavenumbers.
The first four shapes are well known in the standard literature as these interaction terms are the same as in $k$-essence during inflation. The last term is not present in standard $k$-essence scenarios. However, it is nevertheless well known in the context of the decoupling limit of the Galileon theory \cite{Burrage:2010cu} and is peaked in the equilateral shape.

The size of the non-Gaussianity is parameterized by the dimensionless quantity
\begin{align}
    f_\text{NL} = \frac{10}{3} \frac{\sum_m \mathcal{A}_m}{\sum_i k_i^3}~.
\end{align}
Conventionally, to match our results with experiments we are interested in the three different shapes: equilateral, local and orthogonal. We can match our shapes by using the scalar product defined in Ref.~[\citen{DeFelice:2013ar}]. While the second $\mathcal{A}_{\dot{\mathcal{R}}^2\mathcal{R}}$ and third shape $\mathcal{A}_{\mathcal{R}(\partial\mathcal{R})^2}$ are highly correlated with the equilateral shape, the first $\mathcal{A}_{\dot{\mathcal{R}}^3}$ and the last two shapes $\mathcal{A}_{\dot{\mathcal{R}}(\partial\mathcal{R})^2}$, $\mathcal{A}_{\delta L/\delta \mathcal{R}}$ are highly correlated with the local shape.
The current constraints for the three typical shapes/templates are  \cite{Planck:2019kim}
\begin{align}
    f_\text{NL}^\text{local}=-0.9 \pm 5.1, \qquad f_\text{NL}^\text{equilateral} = - 26 \pm 47, \qquad f_\text{NL}^\text{orthogonal}=-38\pm 24~.
\end{align}
We note that in our case the non-Gaussian terms, except for the second-to-last term, scale as $\frac{c_s^2-1}{c_s^2} \, \mathcal{O}(1)$. This behavior is well known from standard $k$-essence models. Therefore, many conventional matter bouncing models with $c_s^2 \rightarrow 0$ are ruled out leading to several no-go theorems for matter bouncing models \cite{Quintin:2015rta,Li:2016xjb}.
However, in our case with $c_s^2 \geq 1$ the non-Gaussianity parameters are of order $\mathcal{O}(1)$ which is compatible with the current experimental constraints. Considering the squeezed limit $k_1 \ll k_2= k_3$ we can match the local non-Gaussianity parameter as
\begin{align}
    f_\text{NL}^\text{local}
    & \simeq - \frac{5}{4} \frac{c_s^2-1}{3 c_s^2} (\epsilon-3 c_s^2) + \frac{5}{12} (\epsilon - 3 c_s^2)
    \nonumber\\
    & = \frac{5}{12 c_s^2} (\epsilon - 3 c_s^2) \nonumber \\
    & = - \frac{5}{4} + \frac{5 }{12 w} \frac{5-n_s}{3 w-1}~.
\end{align}
Using that $w\geq 1$, we obtain $ - \frac{5}{4} \leq f_\text{NL}^\text{loc} \leq - \frac{5}{12}$, which is still inside the current constraints, but in reach for future observations. As for the equilateral and orthogonal shapes, they are still far beyond the reach of current observational constraints.
New observations are needed to further constrain the parameter space.

\section{Discussion}
\label{sec:Discussion}
In this study, we discussed the non-linear interactions for the VCDM model. In general, due to the elliptic equations that eliminate unwanted extra degrees of freedom, it is very challenging to derive the full cubic or higher order interaction. Instead, following Ref.~[\citen{Bartolo:2021wpt}], we focused on the scenario where the energy density of the matter fluid is subdominant, \textit{i.e.}~$\alpha = (\rho + p)/H^2 \ll 1$. This scenario is relevant, either if the background is dominated by the VCDM potential $V(\varphi)$ or if the matter fluid has an equation of state $\rho \simeq - p$, as for dark energy or during inflation.

Modeling the inflaton with a $k$-essence fluid, we showed that, during inflation, we can recover the standard results from GR with $k$-essence as long as the slow-roll conditions for the inflaton hold.
However, if the VCDM potential $V(\phi)$ dominates the background, it is possible to construct non slow-roll inflationary scenarios that can lead to significant local contributions to the bispectrum and violate the Maldacena's consistency relation \cite{Maldacena:2002vr}, despite having a scale-invariant scalar power spectrum, and modes freezing outside the horizon.

In the second part, we applied our results to the bouncing model introduced in Ref.~[\citen{Ganz:2022zgs}]. We showed that the scalar bispectrum is of order $f_\text{NL} = \mathcal{O}(1)$ and compatible with current constraints. Future experiments that would improve the constraints on the tensor-to-scalar ratio or the local non-Gaussianities will be crucial to test this model.

In the future, it would be interesting to explore the possibility to test higher order interactions outside the $\alpha \mathcal{H}^2/k^2 \ll 1$ regime, which was adopted here and in previous works \cite{Bartolo:2021wpt}.

% \acknowledgments
\section*{Acknowledgments}
A.G. received support by the grant No. UMO-2021/40/C/ST9/00015 from the National Science Centre, Poland, and from the DFG under the Emmy-Noether program, project number 496592360.
P.M. acknowledges support by grants from the Research Grants Council of
Hong Kong (Project No. CUHK 14304622 and 14307923), the start-up grant
from the Chinese University of Hong Kong, and the Direct Grant for
Research from the Research Committee of The Chinese University of Hong Kong.
S.M. was supported in part by Japan Society for the Promotion of Science (JSPS) Grants-in-Aid for Scientific Research No.~24K07017 and the World Premier International Research Center Initiative (WPI), MEXT, Japan.
We also thank the editors of the special issue devoted to the memory of V.A. Rubakov for inviting us to contribute with our work.

\appendix

\section{Inflation scenario}
\label{app:Inflation_Toy_Model}
We follow the construction in Ref.~[\citen{Ganz:2022zgs}].
We consider a purely kinetic $k$-essence model with constant equation of state
\begin{align}
    P=P_0 X^{\frac{1+w}{2w}} ~.
\end{align}
We then take a power-law ansatz
\begin{align}
    a \propto t^p~, \qquad a \propto \tau^{\frac{p}{1-p}}~,
\end{align}
with $p>1$ in order to fulfill the inflation requirement $\ddot a>0$. Assuming that $\rho+P>0$ in the time-domain of interest, the VCDM potential can be reconstructed by first solving the equation of motion for the non-dynamical scalar field
\begin{align}
    \phi = \frac{3}{2} \int^\tau \md \tau^\prime a ( \rho + p)~,
\end{align}
and then solve the potential $V(\phi)$ via
\begin{align}
    V = \frac{\phi^2}{3} - \rho~.
\end{align}
The equation of motion for the scalar modes $v_k = z \mathcal{R}_k $ is given by
\begin{align}
      v_k^{\prime\prime} + \left( c_\mathcal{R}^2 k^2 - \frac{z^{\prime\prime}}{z} \right) v_k = 0~.
\end{align}
We assume that, when the modes responsible for the CMB leave the horizon scales, the energy density of the scalar field is highly subdominant, \textit{i.e.} $\alpha \ll 1$, such that we can approximate
\begin{align}
    z^2 \simeq a^2 \frac{\alpha}{c_s^2}, \qquad c_\mathcal{R} \simeq c_s~.
\end{align}
In order to get a scale-invariant scalar power spectrum $n_s = 1$ and frozen scalar perturbations outside the horizon we need to require that $ z \propto \tau^{-1}$, which leads to the condition
\begin{align}
    p = \frac{4}{3(1+w)} >1~, \qquad   w < \frac{1}{3}~.
\end{align}
We further require that $w>0$ to obtain stable scalar perturbations. Note that in the parameter space, $\epsilon$ is of order one
\begin{align}
    1 > \epsilon = \frac{1}{p} = \frac{3(1+w)}{4} > \frac{3}{4}~.
\end{align}
Finally, we obtain the power spectrum
\begin{align}
    P_\mathcal{R} = \frac{H^2}{8 \pi^2 \alpha c_s^2} \big \vert_{c_s k=a H}~.
\end{align}
Lastly, the equation of motion for the tensor modes is given by
\begin{align}
    \left( a h_{ij} \right)^{\prime\prime} + \left( k^2 - \frac{a^{\prime\prime}}{a} \right) (a h_{ij}) = 0~,
\end{align}
leading here to a very red tilted spectrum of index
\begin{align}
    n_t = - 2 \epsilon~.
\end{align}

\bibliographystyle{ws-ijmpa}
\bibliography{bibliography}

\end{document}